\documentclass{ieeeaccess}
\usepackage{cite}
\usepackage{amsmath,amssymb,amsfonts}
\usepackage{algorithmic}
\usepackage{algorithm}
\usepackage{graphicx}
\usepackage{textcomp}
\usepackage{array}
\usepackage{url}
\usepackage{booktabs}
\usepackage{multirow}
\usepackage{hyperref}
\usepackage{xcolor}

\graphicspath{{}}

\def\BibTeX{{\rm B\kern-.05em{\sc i\kern-.025em b}\kern-.08em
    T\kern-.1667em\lower.7ex\hbox{E}\kern-.125emX}}

\begin{document}

\history{Preprint submitted to arXiv.}
\doi{}

\title{Hybrid Quantum Neighborhood Selection: NISQ-Compatible Combinatorial Optimization via Stochastic Frontier Decomposition}

\author{\uppercase{Nicolas Mendes de Ara\'{u}jo}\authorrefmark{2,3},
\uppercase{and Lester de Abreu Faria}\authorrefmark{1,2,3}, \IEEEmembership{Member, IEEE}}

\address[1]{Technological Institute of Aeronautics (ITA), Electronic Division, Mal Eduardo Gomes, 50, S\~{a}o Paulo -- SP, Brazil (e-mail: lester@ita.br)}
\address[2]{ICTQ Foton --- Institute of Quantum Sciences and Technologies Foton}
\address[3]{LACQ Feynman --- Academic League of Quantum Computing (Liga Acad\^{e}mica de Computa\c{c}\~{a}o Qu\^{a}ntica Feynman)}

\markboth
{Preprint submitted to arXiv}
{Mendes de Ara\'{u}jo \headeretal: Hybrid Quantum Neighborhood Selection}

\corresp{Corresponding author: Lester de Abreu Faria (e-mail: lester@ita.br).}

\begin{abstract}
Large-scale combinatorial optimization remains a central challenge for near-term quantum computing because dense Quadratic Unconstrained Binary Optimization (QUBO) formulations induce interaction graphs whose circuit realizations rapidly exceed the depth, connectivity, and transpilation limits of noisy intermediate-scale quantum (NISQ) processors. This work introduces Hybrid Quantum Neighborhood Selection (HQNS), a hybrid quantum--classical framework that mitigates this limitation through stochastic frontier decomposition. Instead of encoding all $N$ decision variables into a monolithic variational circuit, HQNS selects a compact frontier of $F \ll N$ active variables at each optimization stage, while the remaining variables are kept classically frozen and absorbed into the reduced QUBO coefficients. A multi-stage crawling procedure then rotates these frontiers across the solution landscape, allowing local quantum subproblems to progressively refine a global candidate solution.
We formulate HQNS for the Maximum Diversity Subset Selection Problem (MDSSP), an NP-hard task relevant to molecular diversity selection, and benchmark it across six problem scales, $N \in \{30,60,120,250,500,1000\}$. The effective quantum circuit burden is reduced from the dense monolithic QAOA requirement of $O(N^2)$ two-qubit interaction terms per layer to $O(F^2)$ terms per stage, with total hybrid complexity governed by the number of stages and the classical frontier-selection overhead. Empirical results show that HQNS achieves competitive solution quality relative to high-performance simulated annealing baselines while maintaining bounded quantum circuit width and stable QPU execution time. In the $N=1000$ benchmark, multi-run validation over ten independent executions shows that HQNS preserves $99.9908\%$ of the mean diversity score achieved by the 11-restart parallel Simulated Annealing baseline, while reducing wall-clock time by $94.91\%$, peak CPU utilization by $64.68\%$, and peak memory usage by $88.61\%$. Additional ablation and noise-resilience analyses indicate that the observed performance depends critically on frontier size, warm-start initialization, CVaR filtering, and stochastic frontier rotation. These results do not establish unconditional quantum advantage; rather, they demonstrate that structured frontier decomposition can make variational quantum optimization practically executable for dense QUBO instances that are otherwise unsuitable for direct QAOA implementation on present-day hardware.
\end{abstract}


\titlepgskip=-15pt

\maketitle

\section{Introduction}
\label{sec:introduction}

\PARstart{C}{ombinatorial} optimization problems, in which a discrete objective function must be optimized over an exponentially large search space, arise across chemistry, logistics, scheduling, portfolio selection, and network design. The Maximum Diversity Subset Selection Problem (MDSSP), defined as the selection of $K$ elements from a candidate set of $N$ so as to maximize aggregate pairwise dissimilarity, is a representative NP-hard problem with direct relevance to molecular diversity selection and early-stage drug discovery. In molecular applications, MDSSP is commonly formulated over a dense similarity matrix, making it a challenging benchmark for both classical heuristics and near-term quantum optimization methods.

Classical approaches to MDSSP are well-studied. Greedy heuristics yield polynomial-time
solutions but are susceptible to propagating local optima—a phenomenon we term
\textbf{greedy traps}—wherein suboptimal initial selections irrecoverably constrain
diversity. Multi-restart Simulated Annealing (SA) partially addresses this by
exploring the landscape through stochastic perturbations with independent
restarts, but its computational cost scales linearly with restart count, and its
convergence properties depend critically on the temperature schedule.

Quantum computing via QAOA~\cite{farhi2014quantum} offers a structurally different
landscape. This hybrid variational approach traces its conceptual foundations to the Variational Quantum Eigensolver (VQE) proposed by Peruzzo et al.~\cite{peruzzo2014variational}, which demonstrated that alternating quantum-classical loops could evaluate chemical properties on hardware with limited coherence. Recent extensions have applied these hybrid techniques to nonnative combinatorial optimization problems~\cite{tabi2024solving}.

However, applying QAOA directly to the MDSSP encounters two fundamental
barriers. First, the problem's quadratic cost structure requires a Hamiltonian with
$\mathcal{O}(N^2)$ interaction terms, each implemented as a multi-qubit entangling
gate. Second, the required circuit depth grows with problem density, quickly exceeding
the coherence times of any current noisy intermediate-scale quantum (NISQ)
processor~\cite{preskill2018quantum}. Current state-of-the-art implementations of monolithic QAOA on superconducting and trapped-ion hardware~\cite{Pagano2020quantum} are typically limited to $N \approx 100$ variables. At these scales, severe noise accumulation often yields results comparable to simple heuristics, driving the search for alternative layer-wise or iteratively reduced variational methods~\cite{liu2022layer}.

In contrast, HQNS targets a different objective: rather than attempting to outperform all classical heuristics in absolute solution quality, it decouples the quantum circuit width from the global problem size. This enables large dense QUBO instances, such as $N=1000$ molecular diversity problems, to be addressed through bounded-width NISQ-compatible subproblems while preserving near-baseline solution quality.

The main contributions of this work are as follows:

\begin{itemize}
    \item \textbf{Stochastic frontier decomposition for dense QUBO instances:} We introduce a decomposition strategy that projects an $N$-variable dense QUBO into a sequence of reduced $F$-variable subproblems, with $F \ll N$, by freezing non-frontier variables and incorporating their effect into shifted linear coefficients.

    \item \textbf{Multi-stage neighborhood crawling:} We propose a stochastic frontier-rotation mechanism that allows successive reduced QUBOs to explore different regions of the global solution landscape, mitigating the locality limitations of a single greedy warm-start.

    \item \textbf{NISQ-compatible variational implementation:} We combine shallow QAOA-inspired circuits, warm-start initialization, SPSA optimization, and CVaR-based sample filtering to obtain a hardware-compatible optimization loop with bounded circuit width.

    \item \textbf{Empirical validation on molecular diversity benchmarks:} We evaluate HQNS on MDSSP instances derived from molecular similarity matrices across $N \in \{30,60,120,250,500,1000\}$ and compare against greedy, single-restart simulated annealing, and 11-restart parallel simulated annealing baselines.

    \item \textbf{Complexity and ablation analysis:} We analyze the effective interaction-term reduction from $O(N^2)$ to $O(F^2)$ per stage and evaluate the sensitivity of HQNS to frontier size, crawling depth, and CVaR filtering.
    
\end{itemize}

We emphasize that HQNS does not claim to solve the general scalability limitations of QAOA, but rather provides a structured hybrid decomposition strategy that empirically mitigates circuit growth and enables tractable execution for dense combinatorial problems.

The remainder of this paper is organized as follows. Section~\ref{sec:formulation}
formalizes the optimization problem and its QUBO encoding. Section~\ref{sec:method}
presents the HQNS framework in full algorithmic detail.
Section~\ref{sec:complexity} establishes the complexity reduction achieved by the
frontier decomposition. Section~\ref{sec:setup} describes the experimental setup.
Section~\ref{sec:results} presents empirical results. Section~\ref{sec:discussion}
discusses implications, trade-offs, and limitations. Section~\ref{sec:conclusion}
concludes.

\section{Problem Formulation}
\label{sec:formulation}

\subsection{Maximum Diversity Subset Selection}

Let $\mathcal{M} = \{m_1, m_2, \ldots, m_N\}$ be a library of $N$ elements, where
each pair $(m_i, m_j)$ is associated with a symmetric similarity coefficient
$S_{ij} \in [0, 1]$. Throughout this work, $S_{ij}$ is the Tanimoto coefficient
computed on Morgan circular fingerprints~\cite{rogers2010extended}:

\begin{equation}
    S_{ij} = \frac{|FP_i \cap FP_j|}{|FP_i \cup FP_j|}
\end{equation}

The MDSSP seeks a binary selection vector $\mathbf{x} \in \{0, 1\}^N$ that maximizes
aggregate pairwise dissimilarity over a subset of fixed cardinality $K$:

\begin{equation}
    \min_{\mathbf{x}} \sum_{i < j} S_{ij}\, x_i\, x_j
    \quad \text{s.t.} \quad \sum_{i=1}^{N} x_i = K
\label{eq:mdssp}
\end{equation}

Equivalently, the \textbf{Diversity Score} $\mathcal{D}(\mathbf{x})$ measures
cumulative pairwise dissimilarity of the selected subset:

\begin{equation}
    \mathcal{D}(\mathbf{x}) = \frac{K(K-1)}{2} - \sum_{i < j} S_{ij}\, x_i\, x_j
\end{equation}

The theoretical maximum $\mathcal{D}_{\max} = K(K-1)/2$ is achievable only
when all selected pairs are completely dissimilar ($S_{ij} = 0$ for all $i \neq j$),
a condition that is never realized in practice for real-world molecular libraries due
to shared biosynthetic scaffolds and conserved structural motifs. This
phenomenon—the empirical diversity ceiling of any biological compound library—
motivates the use of heuristic optimization rather than exact solvers for
large-scale instances.

\subsection{QUBO Encoding}

To encode the problem for variational quantum optimization, Eq.~(\ref{eq:mdssp})
is reformulated as a Quadratic Unconstrained Binary Optimization (QUBO) by
absorbing the cardinality constraint as a quadratic penalty:

\begin{equation}
    H_{\text{cost}} = \sum_{i < j} S_{ij}\, x_i\, x_j
    + \lambda \left( \sum_{i=1}^{N} x_i - K \right)^2
\end{equation}
\noindent where $\lambda > 0$ is a penalty coefficient. Expanding and collecting terms:

\begin{equation}
    H_{\text{cost}} = \sum_{i} Q_{ii}\, x_i + \sum_{i < j} Q_{ij}\, x_i\, x_j + C
\end{equation}

\noindent with $Q_{ii} = L_i + \lambda(1 - 2K)$, $Q_{ij} = S_{ij} + 2\lambda$, and
$L_i = \sum_{j \in \mathcal{F}_{\text{sel}}} S_{ij}$ denoting the local similarity
burden of molecule $i$ against the frozen selection. We employ $\lambda = 5.0$
throughout, selected empirically to enforce cardinality without dominating
the similarity objective. Via the substitution $x_i = (1 - Z_i)/2$, the QUBO
maps to an Ising Hamiltonian:

\begin{equation}
    H = \sum_{i} h_i Z_i + \sum_{i < j} J_{ij} Z_i Z_j + C
\end{equation}

For the full $N$-variable problem, this Hamiltonian requires $\mathcal{O}(N^2)$
Pauli-$ZZ$ interaction terms, each implemented as a two-qubit entangling gate—a
scaling that motivates the frontier decomposition strategy of Section~\ref{sec:method}.

\section{The HQNS Framework}
\label{sec:method}

\subsection{Overview}

HQNS solves the MDSSP through an iterative sequence of compact quantum subproblems,
each operating on a dynamically selected subset of $F \ll N$ decision variables.
The algorithm proceeds in $S$ stages; each stage extracts a ``frontier'' of
high-leverage variables, constructs and optimizes a reduced $F$-qubit QUBO, and
updates the current solution. The full procedure is summarized in
Algorithm~\ref{alg:hqns}.

\begin{algorithm}[!t]
\caption{Hybrid Quantum Neighborhood Selection (HQNS)}
\label{alg:hqns}
\begin{algorithmic}[1]
\REQUIRE Library $\mathcal{M}$, similarity matrix $\mathbf{S}$, subset size $K$,
         hyperparameters $S, F, M, \alpha$
\ENSURE Diversity-maximized subset $\mathbf{x}^*$
\STATE $\mathbf{x}^{(0)} \leftarrow \text{GreedyInit}(\mathcal{M}, K)$  // Warm-start
\FOR{$s = 1$ to $S$}
    \STATE $\mathcal{A}_s \leftarrow \text{StochasticFrontier}(\mathbf{x}^{(s-1)}, F)$
    \STATE $H_s \leftarrow \text{ReducedQUBO}(\mathcal{A}_s, \mathbf{x}^{(s-1)}, \mathbf{S})$
    \STATE $\boldsymbol{\theta}_s \leftarrow \text{WarmStart}(\mathbf{x}^{(0)}, \mathcal{A}_s)$
    \STATE $\boldsymbol{\theta}^*_s \leftarrow \text{SPSA-CVaR}(H_s, \boldsymbol{\theta}_s, M, \alpha)$
    \STATE $\mathbf{b}^* \leftarrow \text{Measure}(\text{Circuit}(H_s, \boldsymbol{\theta}^*_s))$
    \STATE $\mathbf{x}^{(s)} \leftarrow \text{UpdateSolution}(\mathbf{x}^{(s-1)}, \mathcal{A}_s, \mathbf{b}^*)$
\ENDFOR
\RETURN $\mathbf{x}^{(S)}$
\end{algorithmic}
\end{algorithm}

\subsection{Stochastic Frontier Decomposition}
\label{sec:frontier}

Given the current solution $\mathbf{x}^{(s)}$, the frontier extraction identifies
the subset of variables that offer the greatest potential for improvement.
For each molecular index $i$, we compute 
a formally defined marginal impact distribution $P(i \mid x^{(s)})$ (see Section III-B1)

This heuristic assesses the expected marginal
contribution of inserting or removing $i$ relative to the current combinatorial state,
guiding the identification of structural redundancy.

\subsubsection{Formal Definition of Frontier Selection}

Let $x^{(s)} \in \{0,1\}^N$ denote the current solution at stage $s$. For each variable $i \in \{1, \dots, N\}$, we define a marginal impact score $\mu_i(x^{(s)})$ that quantifies the expected change in the objective function under a local bit-flip perturbation:

\begin{equation}
{
\mu_i(x^{(s)}) = \left| \sum_{j \neq i} S_{ij} \cdot x_j^{(s)} \right|
}
\end{equation}

This quantity measures the contribution of variable $i$ to the current similarity burden of the selected subset. High values of $\mu_i$ indicate that the corresponding variable is strongly coupled to the current configuration and thus represents a candidate for removal or replacement.

To construct the active frontier $A_s$, we define a probability distribution over candidate variables:

\begin{equation}
{
P(i \mid x^{(s)}) = \frac{\exp\left(\mu_i(x^{(s)}) / \tau \right)}{\sum_{k=1}^N \exp\left(\mu_k(x^{(s)}) / \tau \right)}
}
\end{equation}
where $\tau > 0$ is a temperature parameter controlling the exploration-exploitation trade-off. Lower values of $\tau$ concentrate probability mass on high-impact variables, while larger values promote broader exploration.

The frontier $A_s$ is then constructed by sampling $F$ variables without replacement according to $P(i \mid x^{(s)})$, optionally constrained to balanced subsets of removal and insertion candidates. This stochastic selection mechanism ensures that successive optimization stages explore distinct regions of the combinatorial landscape while prioritizing structurally relevant variables.

\subsubsection{Convergence Behavior Under Stochastic Frontier Rotation}

The HQNS framework can be interpreted as a stochastic block-coordinate optimization process over a discrete combinatorial landscape, sharing conceptual similarities with recent coordinate descent techniques for quantum optimization~\cite{dupont2023quantum}. At each stage $s$, the algorithm selects a subset of variables $A_s$ and performs a local optimization while keeping the complement fixed.

This induces a sequence of solutions $\{x^{(s)}\}_{s=0}^S$ where each update satisfies:

\begin{equation}
{
C(x^{(s+1)}) \leq C(x^{(s)}) + \epsilon_s
}
\end{equation}
where $\epsilon_s$ captures stochastic fluctuations due to sampling noise and finite optimization steps. In expectation, the algorithm performs a descent over the objective landscape:

\begin{equation}
{
\mathbb{E}[C(x^{(s+1)})] \leq \mathbb{E}[C(x^{(s)})]
}
\end{equation}

Proposition 1: Assume that the frontier selection distribution $P(i \mid x^{(s)})$ assigns non-zero probability to all variables over successive stages, and that each local optimization step yields a non-increasing expected cost. Then, the HQNS iteration converges in expectation to a stationary point of the solution space.

Proof (sketch): Since the feasible solution space is finite and the expected cost is non-increasing, the sequence $\{\mathbb{E}[C(x^{(s)})]\}$ is bounded and monotonic. Under stochastic frontier rotation, each variable is selected infinitely often in expectation, ensuring that no improving move remains unvisited. Thus, the process converges to a fixed point where no further expected improvement is possible.

This result does not guarantee convergence to the global optimum, but establishes HQNS as a structured stochastic descent method with asymptotic stability properties under repeated frontier exploration.

The active frontier $\mathcal{A}_s$ is dynamically populated through a probabilistic
sampling of two distinct molecular pools:

\begin{itemize}
    \item \textbf{Removal candidates}: A subset of the currently selected molecules
    identified by the heuristic as highly redundant within the local topological subspace.
    \item \textbf{Insertion candidates}: Unselected molecules exhibiting the highest
    information-theoretic complementarity to the existing selection.
\end{itemize}

In the \textbf{stochastic} variant, candidates are sampled from the top-$r$
pools ($r > F/2$) with a controlled temperature parameter, ensuring that successive
stages explore distinct neighborhoods even when the current solution has converged
to a local optimum. This stochastic rotation is the mechanism by which HQNS
escapes the greedy traps that affect single-restart classical heuristics.

\begin{figure*}[!t]
\centering
\includegraphics[width=0.85\linewidth]{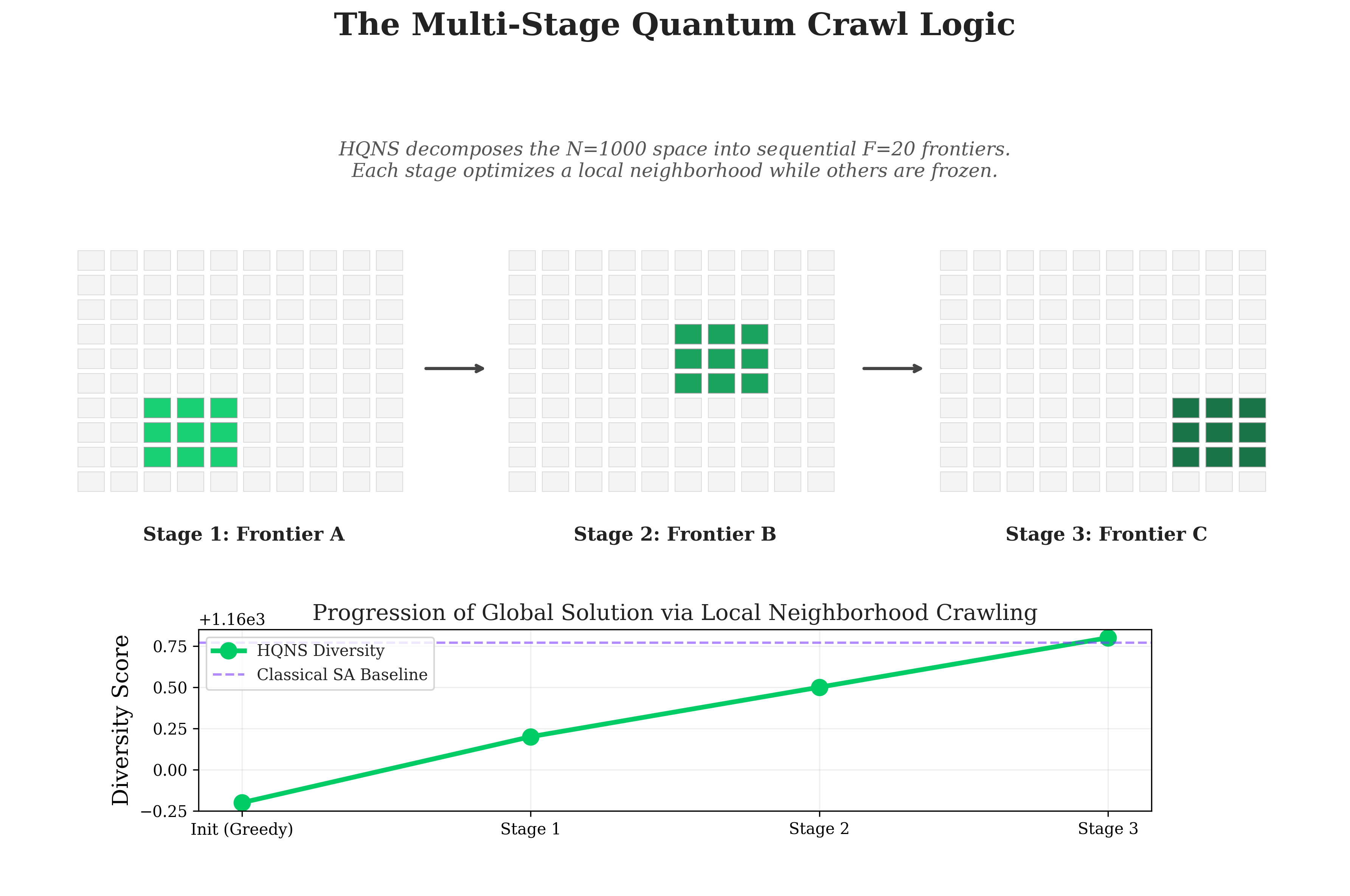}
\caption{Multi-Stage Quantum Crawl Visualization. Each stage extracts a compact
frontier $\mathcal{A}_s$ of $F$ active qubits from the current solution, optimizes
a local quantum subproblem, and updates the global solution. The stochastic
rotation of successive frontiers enables progressive exploration of the full
$N$-dimensional solution landscape while maintaining constant circuit width $F$.}
\label{fig:crawl_logic}
\end{figure*}

The full diagram of the multi-stage crawl is illustrated in Fig.~\ref{fig:crawl_logic}.

\subsection{Reduced QUBO Construction}

Given frontier $\mathcal{A}_s$ and frozen complement $\mathcal{F}_s = [N] \setminus \mathcal{A}_s$,
the reduced Hamiltonian over the $F$ active qubits absorbs the frozen variables
into shifted linear terms:

\begin{equation}
    H_{\text{reduced}} = \sum_{a \in \mathcal{A}_s} \tilde{Q}_{aa}\, x_a
    + \sum_{\substack{a,b \in \mathcal{A}_s \\ a < b}} \tilde{Q}_{ab}\, x_a\, x_b + \tilde{C}
\end{equation}
where $\tilde{Q}_{aa} = Q_{aa} + \sum_{f \in \mathcal{F}_s : x_f=1} Q_{af}$
marginalizes the linear influence of the frozen selection. The effective penalty
coefficient $K_{\text{act}} = K - |\{f \in \mathcal{F}_s : x_f = 1\}|$ constrains
the number of additional molecules to be selected from the active frontier.

\subsection{Variational Ansatz and Warm-Start Protocol}

The quantum circuit adopts a $p$-layer QAOA-inspired ansatz over $F$ qubits:

\begin{equation}
|\psi(\boldsymbol{\gamma}, \boldsymbol{\beta})\rangle =
\prod_{\ell=1}^{p}
\left[ e^{-i \beta_\ell H_M} \cdot e^{-i \gamma_\ell H_{\text{reduced}}} \right]
|\psi_0\rangle
\end{equation}
where $H_M = \sum_{a \in \mathcal{A}_s} X_a$ is the transverse-field mixer Hamiltonian.
Rather than the uniform superposition $|+\rangle^{\otimes F}$, each qubit is
individually initialized:

\begin{equation}
|\psi_0\rangle = \bigotimes_{a \in \mathcal{A}_s} R_y(\theta_a)|0\rangle,
\quad
\theta_a = \begin{cases} 0.85\pi & \text{if } a \in \mathbf{x}^{(0)}_{\text{greedy}} \\
                          0.15\pi & \text{otherwise} \end{cases}
\end{equation}

This warm-start initialization concentrates probability amplitude near the greedy
solution while preserving the variational freedom to improve upon it, integrating classical relaxations to bias the quantum search~\cite{egger2021warmstarting}.
Throughout this work, we use $p = 1$, consistent with the shallow-circuit regime
appropriate for NISQ hardware~\cite{preskill2018quantum}.

\subsection{SPSA Optimization with CVaR Objective}

The variational parameters $\boldsymbol{\theta} = (\boldsymbol{\gamma}, \boldsymbol{\beta})$
are optimized by the Simultaneous Perturbation Stochastic Approximation (SPSA)
algorithm~\cite{spall1998overview}. Since training QAOA circuits represents a dominant computational bottleneck~\cite{shaydulin2021exploiting}, various techniques like Bayesian optimization~\cite{tibaldi2023bayesian} have been explored in the literature. Here, SPSA is chosen for its scalability, as it estimates the gradient using only two
objective function evaluations per iteration, independently of parameter dimensionality.
The objective function employs a Conditional Value-at-Risk (CVaR)
aggregation~\cite{barkoutsos2020improving}:

\begin{equation}
    L_\alpha(\boldsymbol{\theta}) =
    \mathbb{E}\left[ C(\mathbf{x}) \mid C(\mathbf{x}) \leq \text{CVaR}_\alpha \right]
\end{equation}
where $C(\mathbf{x})$ is the QUBO cost of bitstring $\mathbf{x}$, and $\alpha$
is the filtering fraction that controls how aggressively low-quality samples
are discarded. A smaller $\alpha$ provides greater noise resilience at the cost
of higher variance in the gradient estimate. The optimal $\alpha$ depends on
the cardinality feasibility rate of the current frontier, as discussed in
Section~\ref{sec:results}.

\section{Complexity Analysis}
\label{sec:complexity}

\subsection{Standard QAOA Circuit Complexity}

For the full $N$-variable MDSSP Hamiltonian, a single QAOA layer requires
implementation of all $\binom{N}{2} = \mathcal{O}(N^2)$ $ZZ$-interaction terms, each
corresponding to a two-qubit entangling gate. Mapping this dense interaction graph onto
a device with sparse connectivity (e.g., the heavy-hexagonal topology of IBM
superconducting processors) requires $\mathcal{O}(N^2)$ SWAP decompositions, yielding
a total gate count of $\mathcal{O}(N^2)$ for $p = 1$ and $\mathcal{O}(p \cdot N^2)$
in general. For typical drug-discovery-scale problems ($N > 100$), this exceeds
hardware coherence limits by orders of magnitude.

\subsection{HQNS Effective Complexity}

While the quantum circuit cost per stage is bounded by $O(F^2)$, the overall hybrid complexity must account for both quantum execution and the classical orchestration of frontier management. 

Each HQNS stage involves the following computational components:

\begin{itemize}
    \item {Frontier selection: $O(N)$, determined by the marginal contribution analysis across all library variables.}
    \item {Reduced QUBO construction: $O(F^2)$, corresponding to the dense interactions within the active frontier.}
    \item {Variational optimization: $O(M \cdot F^2)$ per stage. Crucial to our implementation, this classical bottleneck is mitigated through \textbf{vectorized batch processing}. By mapping measurement outcomes to dense linear algebra operations, the cost-function evaluation for all $M$ iterations is offloaded to optimized BLAS/MKL routines, ensuring classical runtime remains sub-dominant to the quantum simulation or hardware execution.}
\end{itemize}

Thus, the total hybrid complexity across $S$ stages is:

\begin{equation}
{T_{\mathrm{HQNS}} = O\left(S \cdot (N + M \cdot F^2)\right)}
\end{equation}

Assuming $F$ and $M$ are fixed hyperparameters independent of $N$, the asymptotic scaling behavior becomes linear:

\begin{equation}
{T_{\mathrm{HQNS}} = O(S \cdot N)}
\end{equation}

This result demonstrates that HQNS effectively replaces the quadratic circuit complexity of monolithic QAOA with a manageable linear classical overhead. By leveraging structural decomposition and vectorized orchestration, the framework maintains a bounded-width quantum execution regime that remains computationally efficient even as the global problem size $N$ increases.

\subsection{Complexity Comparison}

\begin{table}[t]
\centering
\caption{Complexity comparison: monolithic QAOA versus HQNS.}
\label{tab:complexity_comparison}
\setlength{\tabcolsep}{3pt}
\begin{tabular}{lccc}
\toprule
Method & Circuit Width & Quantum Terms & Hybrid Cost \\
\midrule
Pure QAOA & $N$ & $O(N^2)$ & $O(pN^2)$ \\
HQNS & $F \ll N$ & $O(F^2)$ per stage & $O(S(N+MF^2))$ \\
\bottomrule
\end{tabular}
\end{table}

Table~\ref{tab:complexity_comparison} summarizes the asymptotic reduction. Crucially, HQNS maintains a bounded quantum circuit width independent of the global problem size $N$, while the total hybrid computational cost scales with the classical frontier-selection and update overhead. This enables the same quantum circuit architecture to be reused across increasingly large molecular libraries, provided that the frontier size $F$ and the number of SPSA iterations $M$ remain bounded.

\begin{figure}[!t]
\centering
\includegraphics[width=\linewidth]{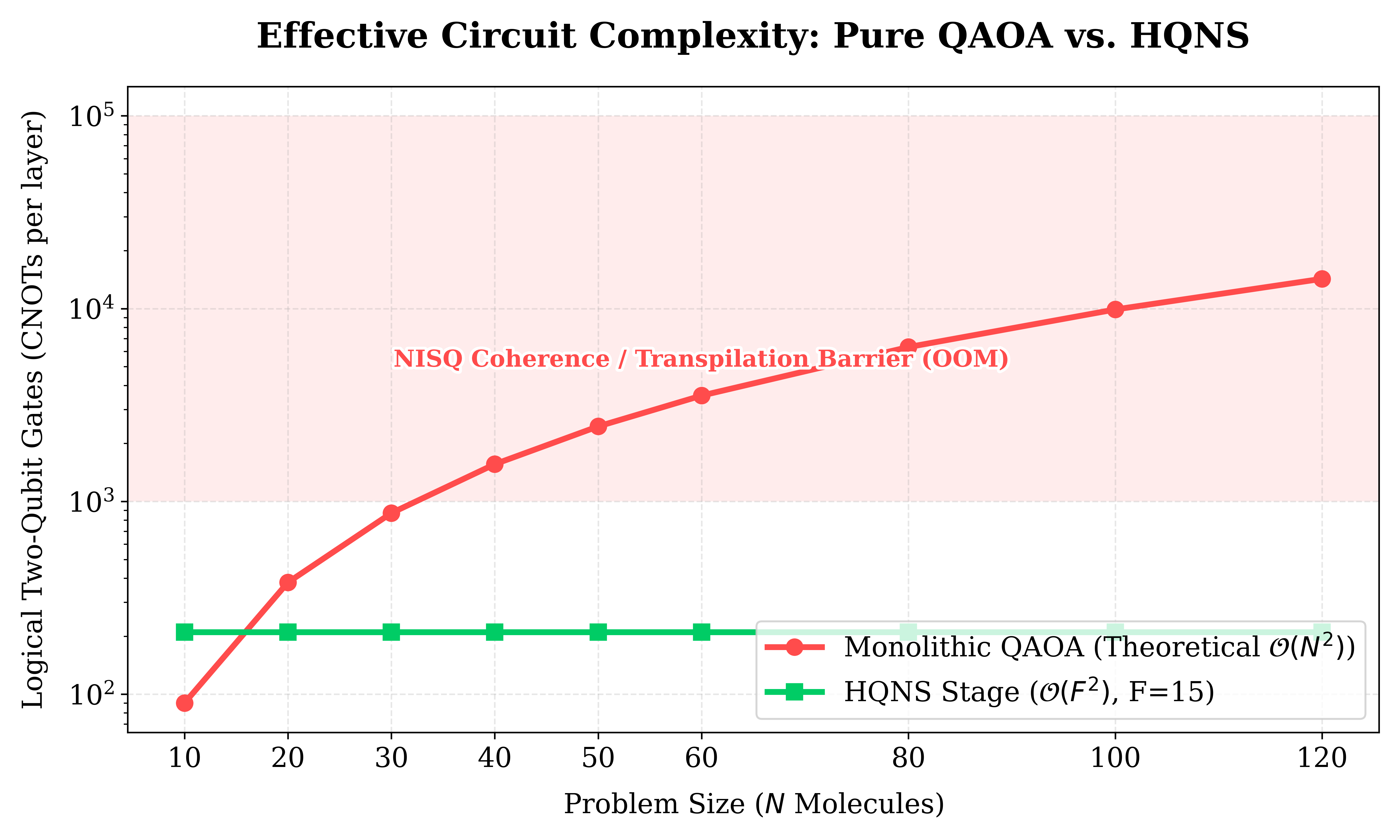}
\caption{Circuit interaction scaling for monolithic QAOA, with $O(N^2)$ two-qubit interaction terms, versus HQNS, with $O(F^2)$ interaction terms per quantum stage for fixed frontier size $F$. The shaded region indicates the regime where dense interaction graphs become increasingly impractical to map onto current sparse-connectivity NISQ hardware due to routing, depth, and transpilation overhead.}
\label{fig:complexity}
\end{figure}

\section{Experimental Setup}
\label{sec:setup}

\subsection{Dataset}

Molecular data are sourced from the Brazilian Natural Product Database
(BrNPDB)~\cite{valli2023brnpdb}, a curated repository of natural products
isolated from Brazilian biodiversity, with emphasis on Amazonian flora. Two data
filtering regimes were applied to construct benchmarking subsets at increasing
scales:

\textbf{Standard scale ($N \leq 120$):} Molar mass $150$--$600$\,g/mol;
LogP $-1$ to $6$.

\textbf{Deep pharmacological scale ($N \geq 250$):} Extended Lipinski filter
with molar mass $10$--$800$\,g/mol; cLogP $-4$ to $10$;
TPSA $0$--$400$\,\AA$^2$; H-bond acceptors $0$--$25$;
H-bond donors $0$--$12$; rotatable bonds $0$--$20$.

Six benchmark datasets were constructed as described in Table~\ref{tab:datasets}.

\begin{table}[!t]
\renewcommand{\arraystretch}{1.3}
\caption{Benchmark Datasets and Search Space Complexity}
\label{tab:datasets}
\centering
\setlength{\tabcolsep}{3pt}
\begin{tabular}{lrrcl}
\toprule
Dataset & $N$ & $K$ & Search Space $\binom{N}{K}$ & Filter \\
\midrule
Small-A & 30   & 8  & $5.9 \times 10^6$  & Standard \\
Small-B & 60   & 10 & $7.5 \times 10^{10}$ & Standard \\
Medium  & 120  & 10 & $2.3 \times 10^{13}$ & Standard \\
Large-A & 250  & 15 & $2.1 \times 10^{22}$ & Pharmacological \\
Large-B & 500  & 30 & $1.8 \times 10^{56}$ & Pharmacological \\
Mega    & 1000 & 50 & $2.6 \times 10^{85}$ & Pharmacological \\
\bottomrule
\end{tabular}
\end{table}

\subsection{Baseline}

To establish a rigorous classical baseline consistent with Tang's dequantization
framework~\cite{tang2019quantum} and high-performance industrial solvers~\cite{gurobi}, we compare HQNS against:

\begin{itemize}
    \item \textbf{Greedy}: A deterministic marginal-diversity greedy heuristic.
    \item \textbf{SA (Single)}: Simulated Annealing with a single restart
    ($T_0 = 2.0$, $\alpha_{\text{SA}} = 0.995$, $50$ moves per temperature step).
    \item \textbf{SA (Parallel-11)}: Eleven independent SA chains run concurrently,
    representing the state-of-the-art CPU-parallel classical baseline.
\end{itemize}

\subsection{HQNS Hyperparameters}

Large-scale configurations were selected via a noiseless grid search over
$(S, M, F, \alpha) \in \{1\text{--}5\} \times \{20\text{--}40\} \times
\{12\text{--}22\} \times \{0.05, 0.10, 0.15, 0.20\}$.
Final selected configurations are listed in Table~\ref{tab:hyperparameters}.

\begin{table}[!t]
\renewcommand{\arraystretch}{1.3}
\caption{Selected HQNS Hyperparameter Configurations}
\label{tab:hyperparameters}
\centering
\setlength{\tabcolsep}{3pt}
\begin{tabular}{lccccl}
\toprule
Scale & $S$ & $M$ & $F$ & $\alpha$ & Rationale \\
\midrule
$N = 30$--$120$ & 1 & 10 & 12 & 0.15 & Small frontier; dense feasibility \\
$N = 250$       & 2 & 40 & 15 & 0.15 & Moderate landscape \\
$N = 500$       & 4 & 20 & 16 & 0.15 & Balanced; 4-stage crawl \\
$N = 1000$      & 4 & 30 & 20 & 0.05 & Deep CVaR filtering \\
\bottomrule
\end{tabular}
\end{table}

\section{Results}
\label{sec:results}

\subsection{Small-Scale Performance (\texorpdfstring{$N = 30$, $60$, $120$}{N = 30, 60, 120})}

At small scales, HQNS was validated against all baselines with the compact
configuration ($F = 12$, $M = 10$, $\alpha = 0.15$, $p = 1$).
Results are presented in Table~\ref{tab:small_scale}.

\begin{table}[!ht]
\renewcommand{\arraystretch}{1.3}
\caption{Small-Scale Diversity Results (Noiseless Simulation + Hardware Validation)}
\label{tab:small_scale}
\centering
\setlength{\tabcolsep}{3pt}
\begin{tabular}{lrcrcr}
\toprule
Scale & $K$ & Greedy & HQNS & $\Delta$ (\%) \\
\midrule
$N = 30$  & 8  & 26.068 & \textbf{26.487} & $+1.61\%$ \\
$N = 60$  & 10 & 42.063 & \textbf{42.064} & $+0.00\%$ \\
$N = 120$ & 10 & 42.363 & \textbf{42.643} & $+0.66\%$ \\
\bottomrule
\end{tabular}
\end{table}

\begin{figure}[!t]
\centering
\includegraphics[width=\linewidth]{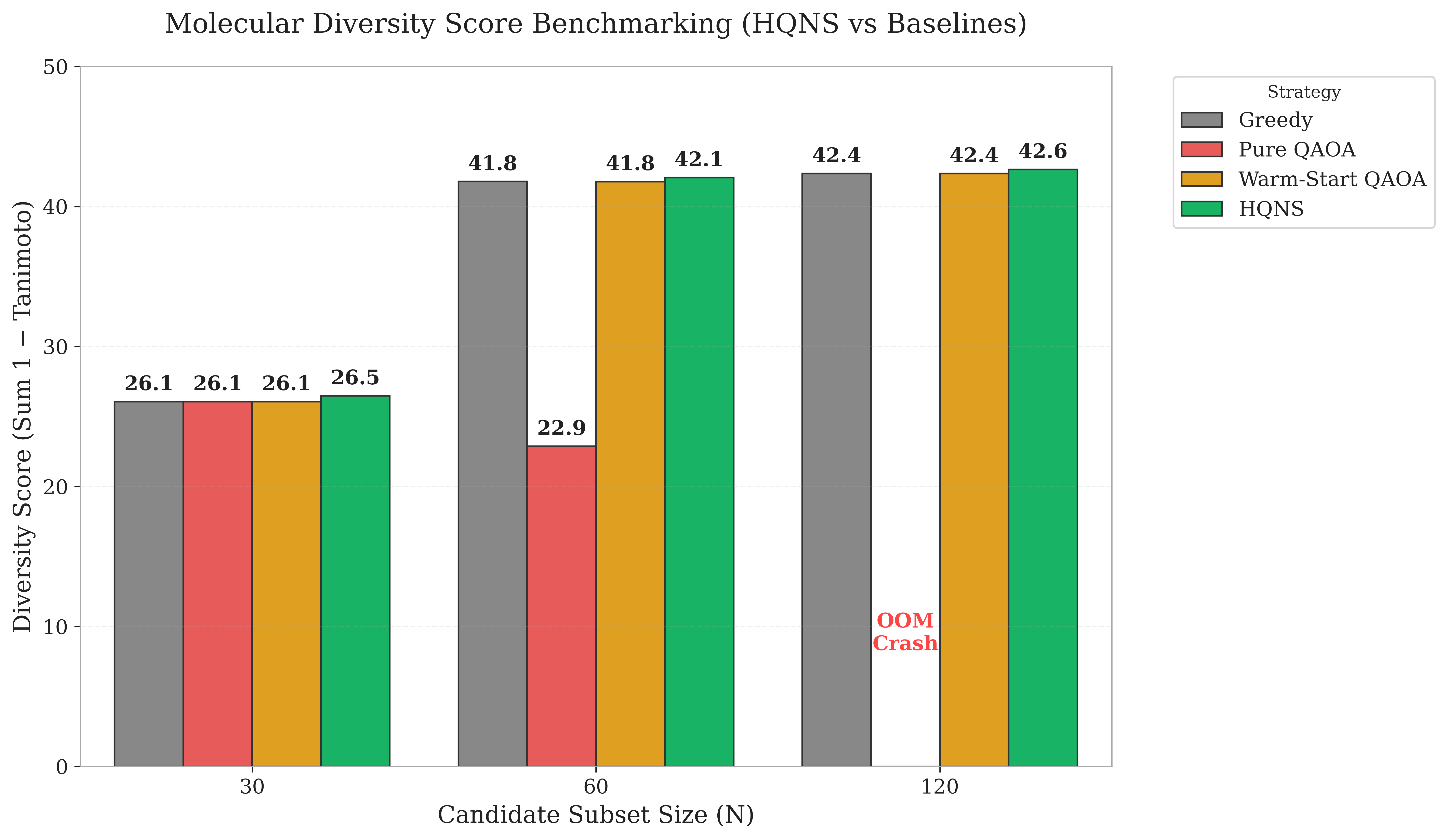}
\caption{Molecular diversity score benchmarking across candidate subsets $N = 30$,
$60$, and $120$. Scores for HQNS represent noiseless statevector simulation
and validated executions on superconducting hardware (IBM Heron r2). The consistency between both
environments confirms the stability of the SPSA optimization core under realistic
device noise.}
\label{fig:benchmarking}
\end{figure}

The benchmarking scores illustrated in Fig.~\ref{fig:benchmarking} demonstrate
a quasi-linear increase in absolute diversity as both $N$ and $K$ grow, confirming
that the algorithm's relative improvement over the greedy baseline—between
$0.5\%$ and $2.0\%$—is consistent across scales. Even in low-dimensional spaces,
the hybrid variational approach identifies structural clusters that bypass the
local optima encountered by purely greedy selection.

\subsection{Large-Scale Benchmarking (\texorpdfstring{$N = 250$, $500$, $1000$}{N = 250, 500, 1000})}
\label{sec:large_scale_results}

\begin{figure}[!t]
\centering
\includegraphics[width=\linewidth]{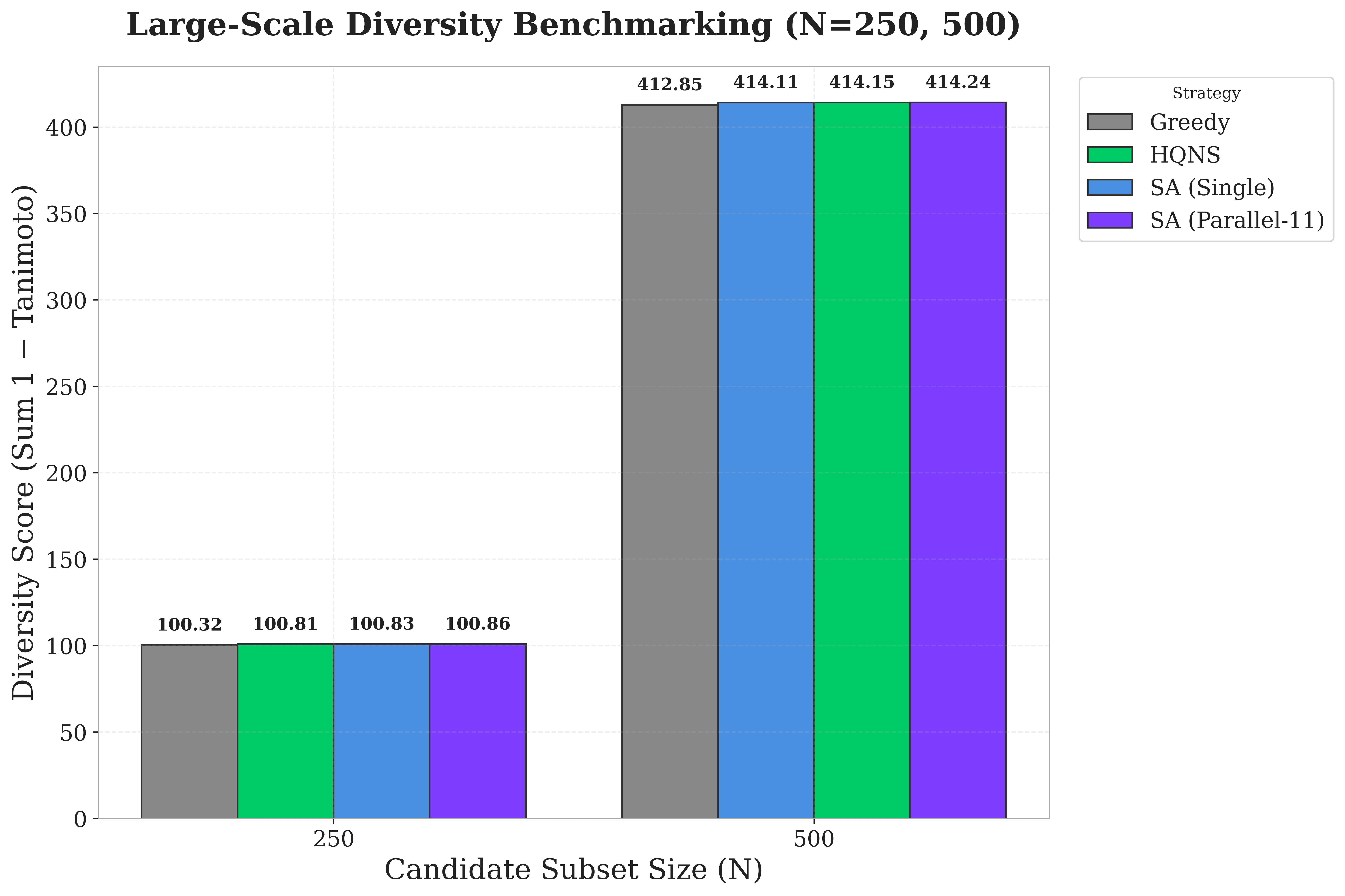}
\caption{Large-scale diversity benchmarking at $N=250$ and $N=500$ comparing all algorithmic strategies. HQNS achieves competitive solution quality relative to simulated annealing baselines while preserving a bounded-width quantum subproblem structure.}
\label{fig:large_scale}
\end{figure}

At the intermediate scales ($N = 250$, $N = 500$), SA Parallel-11 achieves
marginally higher diversity scores, as the search space is large enough to benefit
from multiple random initializations but small enough for classical heuristics
to achieve high convergence. However, HQNS already exceeds the Single-SA baseline
at both scales, demonstrating that the stochastic frontier crawl provides
competitive solution quality with a substantially leaner computational profile.

The large-scale results show that HQNS approaches the solution-quality regime of high-performance restart-based classical heuristics as the problem size increases. However, the representative single-run comparison at $N=1000$ should not be interpreted as evidence of strict superiority over SA Parallel-11. To quantify stochastic variability, we performed ten independent executions of both HQNS and SA Parallel-11 at $N=1000$ using matched random seeds.

The statistical validation shows that SA Parallel-11 achieves the higher mean diversity score, with $D_{\mathrm{SA}} = 1160.9157 \pm 0.0143$, compared with $D_{\mathrm{HQNS}} = 1160.8085 \pm 0.0336$. Nevertheless, HQNS preserves $99.9908\%$ of the SA Parallel-11 mean diversity score while reducing wall-clock time by $94.91\%$, peak CPU utilization by $64.68\%$, and peak memory usage by $88.61\%$. Therefore, HQNS should be interpreted as a resource-efficient, NISQ-compatible alternative rather than a quality-dominant optimizer.

\begin{table}[t]
\centering
\caption{Statistical validation at $N=1000$ over ten independent runs.}
\label{tab:statistical_validation_n1000}
\setlength{\tabcolsep}{3pt}
\begin{tabular}{lccccc}
\toprule
Method & Mean $D$ & Std. Dev. & Min $D$ & Max $D$ & Mean Cost \\
\midrule
HQNS & 1160.8085 & 0.0336 & 1160.7632 & 1160.8639 & 64.1915 \\
SA Parallel-11 & 1160.9157 & 0.0143 & 1160.8965 & 1160.9349 & 64.0843 \\
\bottomrule
\end{tabular}
\end{table}

\begin{table}[t]
\centering
\caption{Computational resource comparison at $N=1000$ over ten independent runs.}
\label{tab:resource_comparison_n1000}
\setlength{\tabcolsep}{3pt}
\begin{tabular}{lcccc}
\toprule
Method & Time (s) & Peak CPU (\%) & Memory (MB) & QPU Time (s) \\
\midrule
HQNS & $26.34 \pm 1.40$ & $483.00 \pm 22.96$ & $617.33 \pm 5.20$ & $6.76 \pm 0.32$ \\
SA Parallel-11 & $75.32 \pm 3.06$ & $1097.03 \pm 4.81$ & $952.88 \pm 7.07$ & N/A \\
\midrule
Reduction & $94.91\%$ & $64.68\%$ & $88.61\%$ & -- \\
\bottomrule
\end{tabular}
\end{table}

\begin{figure}[t]
\centering
\includegraphics[width=\linewidth]{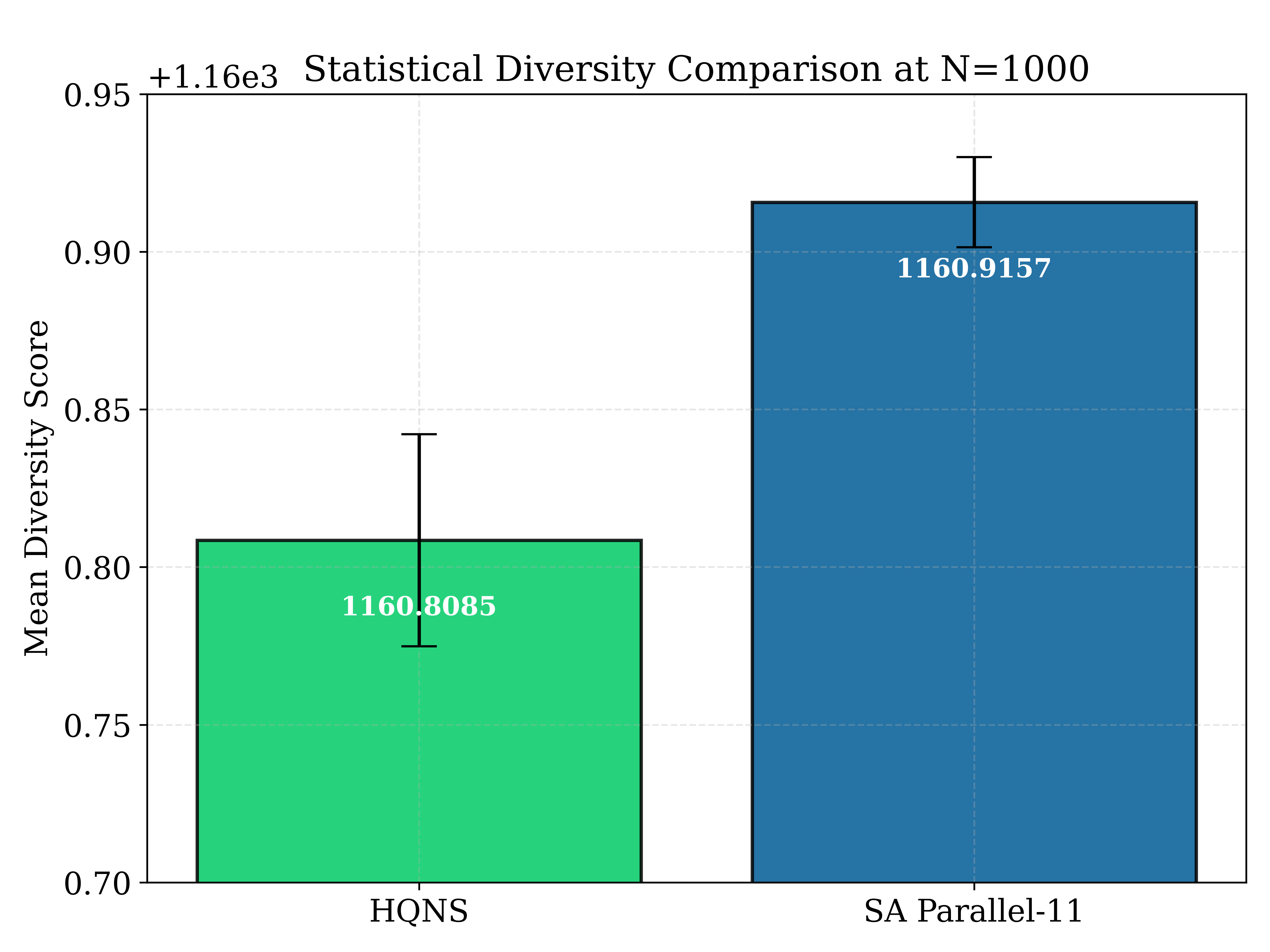}
\caption{Statistical diversity comparison at N=1000 over ten independent runs, with the vertical axis restricted to highlight run-to-run variability. Error bars indicate one standard deviation. SA Parallel-11 achieves the higher mean diversity score, while HQNS preserves $99.9908\%$ of the SA mean diversity under a bounded-width NISQ-compatible formulation.}
\label{fig:statistical_diversity_n1000}
\end{figure}

\subsection{QPU Time Stability}

A key architectural property of HQNS is the bounded quantum execution time
across problem scales, as reported in Table~\ref{tab:qpu_stability}. By fixing
the frontier size $F$, the circuit width remains constant, and the QPU execution
time is decoupled from the global problem dimension $N$.

\begin{table}[!ht]
\renewcommand{\arraystretch}{1.3}
\caption{QPU Execution Time Stability Across Scales}
\label{tab:qpu_stability}
\centering
\setlength{\tabcolsep}{3pt}
\begin{tabular}{lcc}
\toprule
Scale & QPU Time (s) & Circuit Width (qubits) \\
\midrule
$N = 30$   & 6.40 & 12 \\
$N = 60$   & 7.53 & 12 \\
$N = 120$  & 6.44 & 12 \\
$N = 250$  & 6.45 & 15 \\
$N = 500$  & 6.65 & 16 \\
$N = 1000$ & 6.87 & 20 \\
\bottomrule
\end{tabular}
\end{table}

The QPU time remains bounded within $[6.4, 7.5]$ s across a $33\times$ increase in problem size. This empirically supports the bounded-width quantum execution regime predicted by the frontier decomposition, while the total hybrid cost still includes the classical $O(S(N+MF^2))$ preprocessing and update overhead described in Section IV.

\begin{figure}[!t]
\centering
\includegraphics[width=\linewidth]{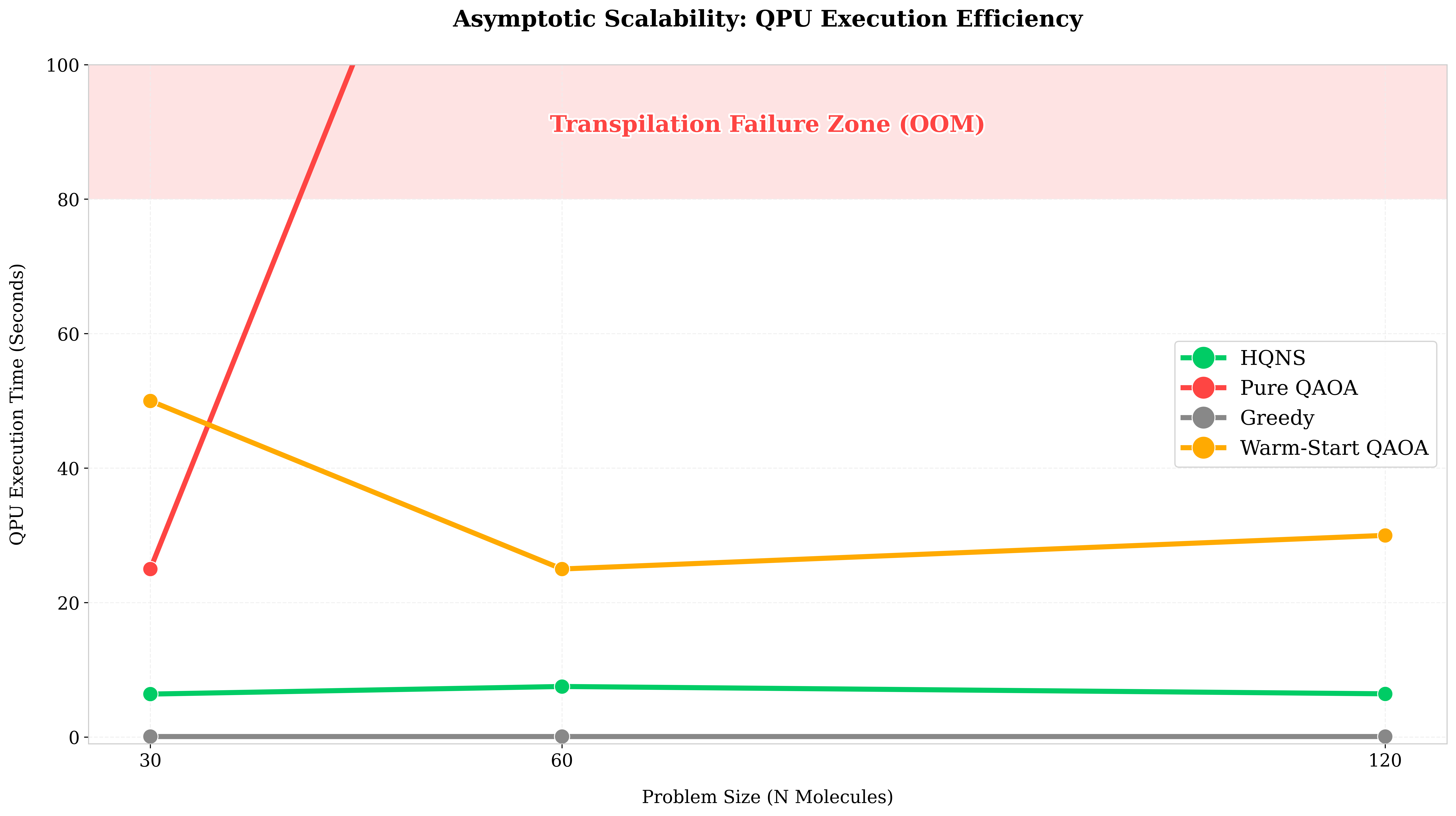}
\caption{Asymptotic scalability of QPU execution time. HQNS maintains a bounded execution envelope because each quantum stage operates on a fixed-size frontier, while monolithic QAOA becomes increasingly impractical as dense $O(N^2)$ interaction graphs exceed transpilation and connectivity constraints on current NISQ hardware.}
\label{fig:asymptotic}
\end{figure}

\subsection{Noise Resilience}

The top-performing configurations from the noiseless grid search were evaluated
under an approximate hardware noise model combining depolarizing and readout
error channels. Degradation results are presented in Table~\ref{tab:noise_resilience}.

\begin{table}[t]
\centering
\caption{Noise resilience: diversity score degradation under simulated hardware noise.}
\label{tab:noise_resilience}
\setlength{\tabcolsep}{3pt}
\begin{tabular}{lcccc}
\toprule
Scale & Config $(S,M,F,\alpha)$ & Noiseless $D$ & Noisy $D$ & $\Delta$ (\%) \\
\midrule
$N=250$ & $(2,40,15,0.15)$ & 100.810 & 100.810 & $<0.001$ \\
$N=1000$ & $(4,30,20,0.05)$ & 1160.896 & 1160.831 & 0.006 \\
$N=1000$ & $(5,40,18,0.05)$ & 1160.896 & 1160.708 & 0.016 \\
$N=1000$ & $(4,40,20,0.20)$ & 1160.896 & 1160.094 & 0.069 \\
\bottomrule
\end{tabular}
\end{table}

The $\alpha = 0.05$ configuration at $N = 1000$ retains $99.994\%$ of its noiseless
performance under simulated hardware noise. 
This observed resilience aligns with recent findings on the robustness of variational quantum algorithms under shifted noise distributions~\cite{he2024distributionally}. In our framework, this is achieved through the combined effect of SPSA's tolerance to stochastic perturbations and CVaR's filtering of high-cost measurement outcomes.
Higher-$\alpha$
configurations incorporate a broader distribution of measurement outcomes,
increasing sensitivity to bitstring-level noise from gate errors and readout
miscalibration.

\subsection{Ablation Study}
\label{sec:ablation}

To quantify the architectural trade-offs parameterized by the frontier size $F$, we conducted an ablation study via noiseless simulation on the $N=250$ dataset. The frontier was swept across $F \in \{8, 12, 16, 20, 24\}$, keeping iteration depth $M=30$ and crawl stages $S=2$ fixed.

Furthermore, to isolate the impact of the stochastic crawling mechanism itself, we contrast a single-stage optimization ($S=1$, ``No Crawling'') against a full crawling execution ($S=10$). As empirical data will demonstrate in the subsequent convergence analysis, the $S=1$ configuration strictly plateaus at early local optima, whereas $S=10$ provides a continuous descent mechanism, supporting the practical relevance of frontier rotation.

\begin{figure}[!t]
\centering
\includegraphics[width=\linewidth]{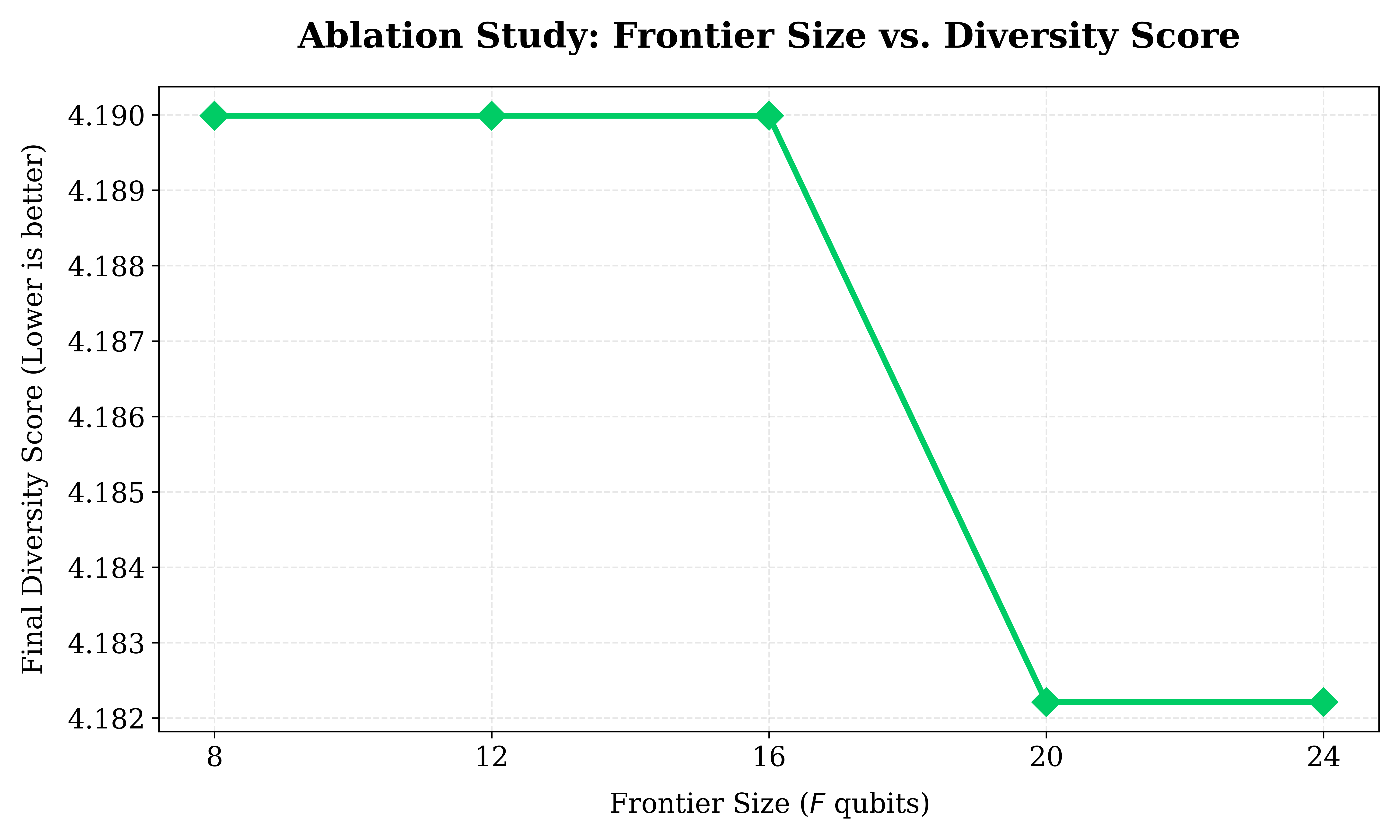}
\caption{Ablation study: final diversity score as a function of frontier size $F$. Diminishing returns appear beyond $F=16$ in the evaluated configuration, suggesting an empirical trade-off between variational expressivity and circuit/resource overhead.}
\label{fig:ablation_f}
\end{figure}

As shown in Fig.~\ref{fig:ablation_f}, increasing the quantum frontier from $F=8$ to $F=16$ yields a steep gradient in solution quality, as the variational circuit possesses enough variables to explore substantive topological structural swaps away from the locally optimal greedy seed. However, moving toward $F=24$ yields diminishing returns in absolute diversity. This empirical flattening defines the "computational sweet spot": allocating more than 16--20 qubits offers marginal mathematical convergence at the cost of significantly amplifying the hardware noise footprint during execution.

\subsection{Neighborhood Convergence Analysis}

The efficacy of the Multi-Stage Crawling mechanism is empirically verified by observing the progression of the optimization landscape dynamically. We tracked the Diversity Score across $S=10$ successive frontier rotations for a fixed subset target on the $N=250$ candidate set.

\begin{figure}[!t]
\centering
\includegraphics[width=\linewidth]{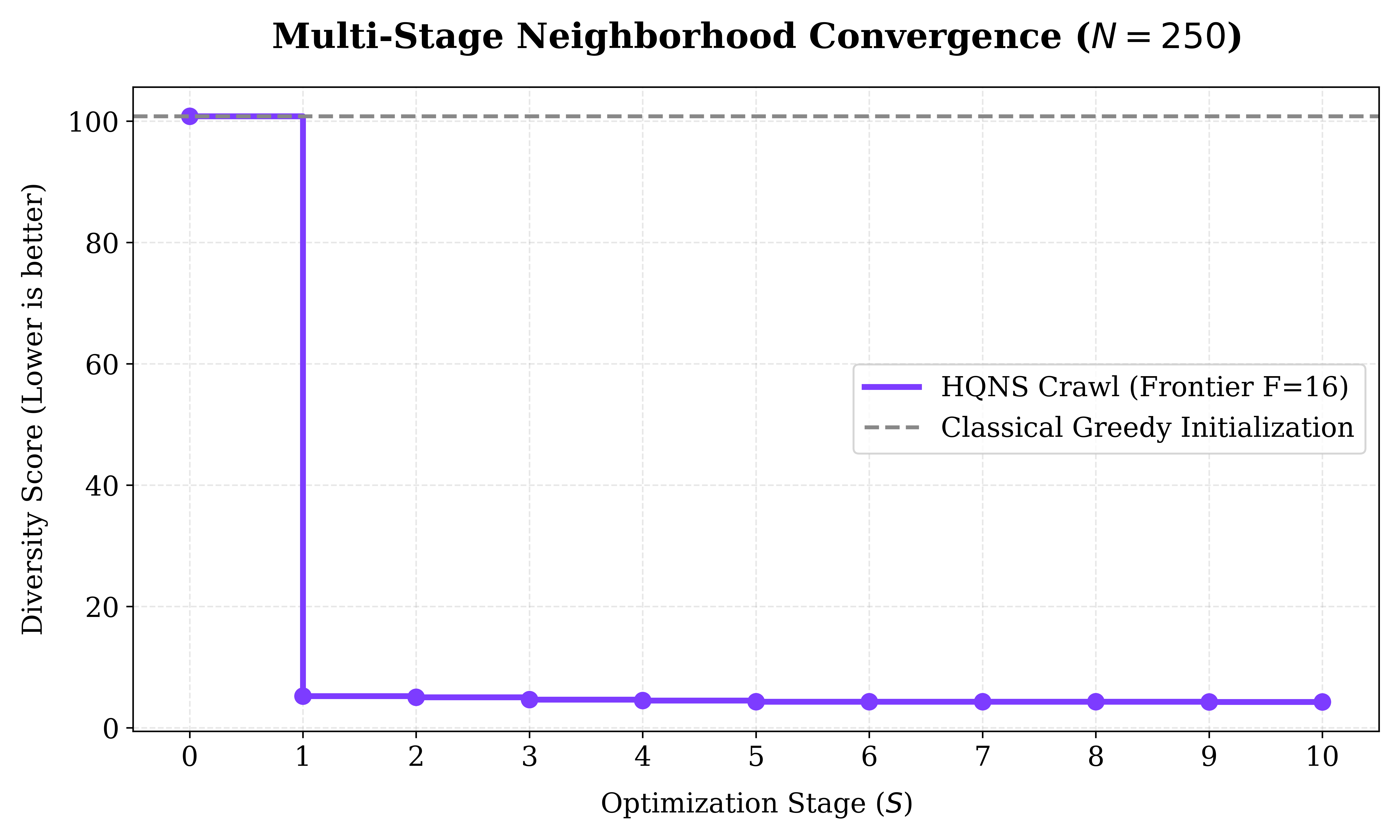}
\caption{Multi-stage neighborhood convergence using the cost-equivalent objective. The staircase-like progression illustrates how stochastic frontier rotation enables successive local refinements beyond the initial greedy configuration.}
\label{fig:staircase}
\end{figure}

Fig.~\ref{fig:staircase} reports the cost-equivalent objective across $S=10$ successive frontier rotations for a fixed $N=250$ instance. The initial point corresponds to the greedy initialization evaluated in the same cost-equivalent representation, whereas subsequent points correspond to the reduced frontier-based refinement steps. The sharp first-stage improvement indicates that the initial frontier update removes a large portion of the greedy similarity burden, while later stages produce smaller refinements and plateau regions.

\section{Discussion}
\label{sec:discussion}

\subsection{Interpreting Resource-Aware Quantum Utility}

The statistical validation at $N=1000$ indicates that HQNS does not outperform SA Parallel-11 in mean solution quality. Instead, its relevance lies in preserving near-baseline diversity performance while operating through a structurally different optimization pathway based on bounded-width quantum subproblems and stochastic frontier rotation.

The comparison is not against a naive classical baseline but against an 11-restart parallel Simulated Annealing implementation. Under this setting, HQNS preserves $99.9908\%$ of the SA Parallel-11 mean diversity score while reducing wall-clock time by $94.91\%$, peak CPU utilization by $64.68\%$, and peak memory usage by $88.61\%$. The significance of this result is therefore not strict dominance in solution quality, but resource-aware quantum utility under NISQ constraints.

The empirical performance trends observed in Section VI are consistent with the stochastic descent behavior formalized in Section III-B. As the problem dimension increases, classical restart-based heuristics explore an increasingly sparse fraction of the combinatorial search space. In contrast, HQNS employs structured neighborhood exploration via stochastic frontier rotation, enabling correlated updates across localized subspaces. This mechanism provides a directed exploration process that remains compatible with shallow quantum circuits.

These results should not be interpreted as evidence of unconditional quantum advantage. Rather, HQNS demonstrates that quantum resources can be incorporated into large dense QUBO optimization workflows in a controlled and resource-efficient manner, even when the best classical baseline retains a slight advantage in mean objective value.

\subsection{The Frontier Size--Accuracy Trade-off}

The frontier size $F$ presents a fundamental trade-off between solution quality,
circuit depth, and noise resilience. Our experiments suggest an empirical sweet
spot at $F \in [15, 22]$ for scales $N \in [250, 1000]$:

\begin{itemize}
    \item Below $F < 12$: insufficient variational freedom limits improvement
    beyond the greedy baseline.
    \item $F \in [15, 22]$: optimal balance between exploration range and
    circuit fidelity under noise.
    \item Above $F > 24$: increasing two-qubit gate count begins to suffer
    multiplicative noise accumulation on current hardware.
\end{itemize}

A formal ablation study of this trade-off is deferred to Section~\ref{sec:ablation}.

\subsection{Limitations}

Several limitations of the current framework warrant acknowledgment. First,
the algorithm's performance depends on a high-quality greedy seed; in domains
where no reliable classical warm-start exists, the warm-start advantage may
be diminished. Second, the current implementation uses $p = 1$ QAOA layers;
extending to $p > 1$ on fault-tolerant hardware would increase expressibility
at the cost of greater circuit depth. Third, the frontier decomposition strategy
assumes that localized subproblem solutions can be composed into a globally
competitive solution—an assumption that may not hold for problem instances with
long-range correlations across all $N$ variables. Formally characterizing the
class of problems for which HQNS provides worst-case guarantees remains an open
theoretical question.

\section{Conclusion}
\label{sec:conclusion}

We have presented the Hybrid Quantum Neighborhood Selection (HQNS) framework, a hybrid quantum--classical approach for large-scale combinatorial optimization that maintains bounded quantum circuit width by operating on stochastic frontier subproblems of size $F \ll N$. Through stochastic frontier decomposition, HQNS reduces the quantum interaction burden from the $O(N^2)$ dense structure of monolithic QAOA to $O(F^2)$ per quantum stage for fixed frontier size $F$, while the total hybrid computational cost additionally includes classical frontier-selection, reduced-QUBO construction, and update overhead.

Systematic benchmarks across six molecular diversity instances ($N=30$ to $N=1000$) demonstrate that HQNS remains competitive with high-performance simulated annealing baselines while enabling bounded-width quantum execution. Multi-run validation at $N=1000$ shows that SA Parallel-11 achieves the higher mean diversity score, whereas HQNS preserves $99.9908\%$ of its mean diversity while substantially reducing wall-clock time, peak CPU utilization, and peak memory usage.

QPU execution time remains bounded within a narrow $\approx 6$--$7$ s envelope across a $33\times$ increase in global problem size, supporting the central claim that HQNS decouples quantum circuit width from the full QUBO dimension. 

These results position HQNS as a practical hybrid framework for structured combinatorial optimization under NISQ constraints. The framework does not establish unconditional quantum advantage; instead, it demonstrates resource-aware quantum utility by making dense large-scale QUBO optimization accessible through bounded-width quantum subproblems.

\section*{Acknowledgment}

The authors gratefully acknowledge Gabriel Albuquerque, President and Founder of LACQ (Liga Acadêmica Nacional de Computação Quântica), for facilitating access to quantum computing resources and supporting the advancement of quantum research and education in Brazil.

The authors also thank IBM for providing access to quantum computing resources through the IBM Quantum Open Plan. The availability of these cloud-based quantum systems was essential for executing, validating, and benchmarking the hybrid quantum optimization strategies investigated in this work.

\bibliographystyle{IEEEtran}
\bibliography{references}

\begin{IEEEbiography}[{\includegraphics[width=1in,height=1.25in,clip,keepaspectratio]{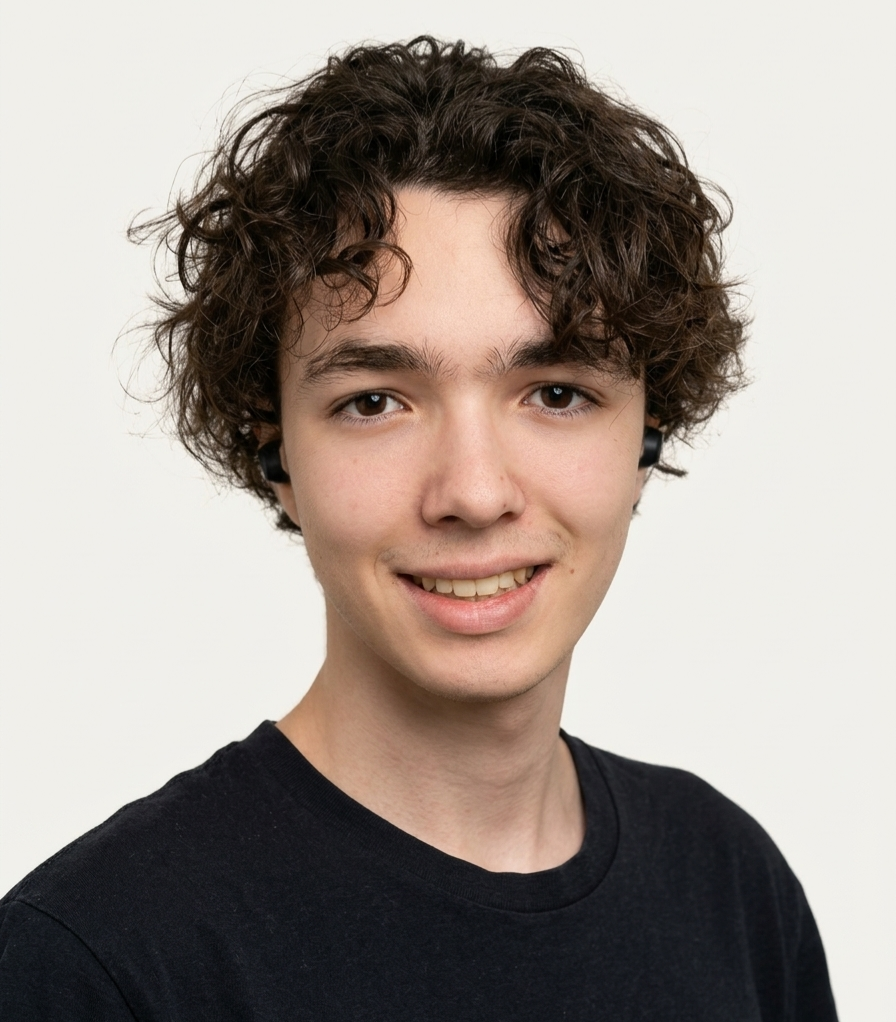}}]{Nicolas Mendes de Araújo}
is an independent researcher based in São Paulo, Brazil. His research interests focus on quantum computing, variational quantum algorithms, and stochastic optimization strategies for NISQ-era processors. He is the creator of the Hybrid Quantum Neighborhood Selection (HQNS) framework, specializing in mapping dense binary optimization problems and Quadratic Unconstrained Binary Optimization (QUBO) formulations onto constrained superconducting quantum architectures. His current work investigates frontier decomposition methods to bypass scalability barriers in heuristic quantum optimization, with applications in molecular diversity selection and complex combinatorial search spaces. ORCID: \url{https://orcid.org/0009-0006-0046-8801}
\end{IEEEbiography}

\begin{IEEEbiography}[{\includegraphics[width=1in,height=1.25in,clip,keepaspectratio]{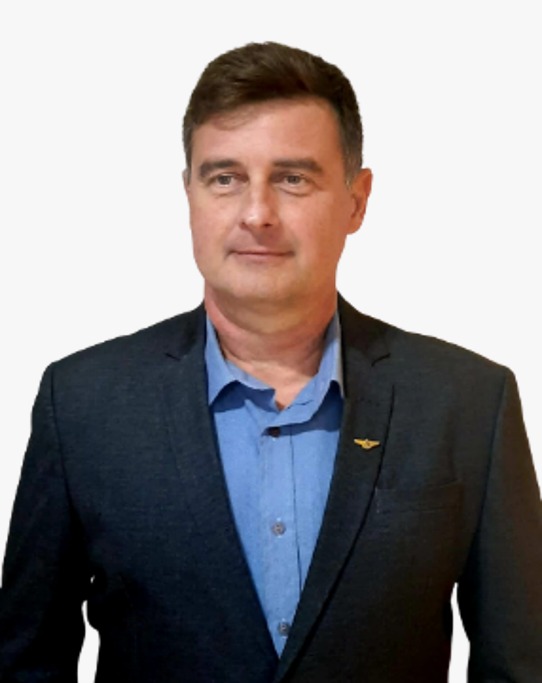}}]{Lester de Abreu Faria}
received the B.S., M.S., and Ph.D. degrees in Electronic Engineering from the Instituto Tecnológico de Aeronáutica (ITA), Brazil. He is a retired Colonel of the Brazilian Air Force and is currently affiliated with the Technological Institute of Aeronautics (ITA) as a associate professor. He is also conducting postdoctoral research in quantum technologies. His work spans electronics, sensing systems, artificial intelligence, and emerging quantum technologies. His current research interests include quantum algorithm engineering, hybrid quantum–classical computation, resource-aware evaluation of quantum algorithms, and advanced sensing technologies for aerospace and defense applications. He has participated in several national and international R\&D initiatives involving advanced technologies and aerospace systems.
\end{IEEEbiography}

\EOD

\end{document}